\documentclass[11pt,eps]{article}

\usepackage{graphicx}
\usepackage{epstopdf}
\begin{document}

 \begin{center}
 
 {Preprint IBR-TP-06-39, November 14, 2006, sumitted to {\it Physics Today}}
 \vskip1.00cm

  {\large {\bf  INSUFFICIENCIES OF THE 20-TH CENTURY THEORIES 
     AND THEIR NEGATIVE ENVIRONMENTAL IMPLICATIONS}}
 \vskip0.30cm

{\bf Ruggero Maria Santilli}

  {President}

 { The Institute for Basic Research}

  {P. O. Box 1577, Palm Harbor, FL 34683, U.S.A.}

{CV at http://www.i-b-r.org/Ruggero-Maria-Santilli.htm}

\begin{abstract}
Following a courageous denunciation by Barton Richter  in {\it Physics Today,} November 2006, of contemporary particle physics as being "theological speculations," we present insufficiencies of special relativity, quantum mechanics, superconductivity, chemistry, standard model, gravitation and astrophysics for conditions beyond those of their original conception.
\end{abstract}

\end{center}
\vskip0.50cm

\noindent {\bf FOREWORD}

   \noindent The author would like to express his sincere appreciation to Burton Richter for his   
  courageous denounciation, in {\it Physics Today} of November 2006, of the current condition of particle physics as  being esentially that of  "theological speculations."
  
  At the same time the author would like to lament that Richter did not enter with sufficient technical 
  depth into the ultimate origin of the ongoing crisis in particle physics because, 
  in so doing, he becomes part of the problem.
  
  As well known, our planet is afflicted by increasingly cataclysmic climactic changes whose solution 
  requires new clean energies and fuels. But all industrially meaningful new energies and fuels 
  based on conventional disciplines had been 
  discovered by the middle of the 20-th century.
  
  Hence, the solution of said environmental problems crucially depends on the identification of 
  the conditions of unequivocal applicability of the doctrines of the 20-th century, as well as 
  the broader conditions under which they are at best approximately valid, thus permitting the 
  search of broader vistas for much needed basic advances. 
  \vskip0.50cm

\noindent {\bf INSUFFICIENCIES OF SPECIAL RELATIVITY}

\noindent The author has often indicated in his papers that Special Relativity (SR) 
has a majestic mathematical structure as well as historical verifications for the 
conditions originally conceived by Einstein and limpidly identified 
in his writing, namely, for point-like particles 
and electromagnetic waves propagating in vacuum, thus including relativistic 
treatments of the hydrogen atom, particle accelerators, and similar systems.

Admittedly, doubts on the final character of SR within the arena of its original 
conception have been voiced for about one century, and have lately increased 
particularly in view of what is becoming the ultimate scientific frontier of 
basic research, the conception of space as a universal medium.

Nevertheless, SR works well for systems verifying its original conditions, such as for  
particle accelerators. Also, these conditions have no known relevance for the 
solution of current environmental problems. Consequently, we believe that the first 
reason for the  ongoing  "theological speculations" is the widespread lack of attention 
on the limitations of SR for physical conditions beyond these of its 
original conception and experimental verification,  as outlined below.
\vskip0.30cm

{\it ANTIMATTER:}  SR (as well as General Relativity, see below) provides no meaningful 
classical representation of antimatter, 
in which field it should be considered "inapplicable," rather than "violated," 
because antimatter had yet to be discovered at the time of conception of the theory.

This limitation is established by the absence of any classical 
distinctions  between neutral systems of matter and antimatter (that are particularly important for stars). 
Even for the case of charged particles it is easy to see the inconsistency of the  
classical treatment (essentially done via the sole change of the sign of the charge) 
because, e.g., the operator image of such a representation is given by a "particle" 
with the wrong sign of the charge, rather than a charge conjugated antiparticle.

Even though considered innocuous by a number of colleagues, in reality the above limitation 
carries deep implications for numerous segments of contemporary physics 
including "theological" (rather than scientific) studies on cancer treatments via 
antiparticles, the gravity of elementary antiparticles in the field of matter, 
not to mention catastrophic inconsistencies in grand unified theories 
without due attention to the inclusion of antimatter.
\vskip0.30cm

{\it NONLOCALITY:}  For the evident purpose of enlarging the arena of applicability of SR, 
the physical universe was generally believed during the 20-th century to be reducible to 
point-like particles that, having no dimension, would always move in vacuum. 

However, by the second part of the 20-th century it became already clear that 
such a view is a mere approximation due to the fact that particles are generally 
extended and, even when having point charges (such as the electron) they have indeed a 
 extended wavepackets causing  
nonlocal interaction of  integral  type (because extended over the volume of mutual 
penetration) that, as such, cannot be reduced to a finite number of points, as needed for SR 
to apply.

Nonlocal interactions of the above type are dramatically beyond 
any possibility of scientific representation by SR for numerous reasons, such as: the basic topologic 
of SR is inapplicable being local-differential, thus implying the loss of Lie's theory 
with consequential loss of the Lorentz and Poincare' symmetries; contact interactions 
are not derivable from a Lagrangian or a Hamiltonian, while being of "zero range" 
or of instantaneous character; the nonlocal interactions here considered are 
generally nonlinear in the wavefunctions; and other reasons.

For "theological" purposes, the nonlocal-integral interactions caused by deep wave 
overlapping are generally dismissed as having no valid physical effects. However,  
in the last part of the 20-th century these 
interactions  came to full light in chemistry because, after 
one century of failed efforts, it became impossible to achieve an exact representation 
of molecular binding energies (due to the historical missing of 2 \%). 

In turn, this 
limitation forced departures from quantum axioms via the introduction of the 
so-called "screened Coulomb potentials" (see below for the loss of the very quantum of 
energy under these adulteration of fundamental axioms). By contrast, recent studies have  
shows that an exact representation of molecular characteristics emerges 
when valence electrons are represented as they are in realty, point-like charges with 
extended wavepackets in conditions of deep mutual penetration, 
thus experiencing not only potential interactions of Coulomb type, but also 
contact nonlocal, nonlinear and non-Hamiltonian interactions.

The implications for particle physics of the above historical occurrence in chemistry begin to 
illustrate rather dramatically the "theological" nature of most contemporary particle research.  
If nonlocal, nonlinear and non-Hamiltonian interactions have such  implications 
in the deep wave overlappings of valence electrons, they are expected to have essentially 
similar numerical implications for deep inelastic scattering of particles. 

But  
the totality of contemporary particle experiments in   
inelastic scatterings are conducted via the century old "potential scattering theory" that 
simply cannot represent nonlocal-integral efforts expected to be inevitable for deep 
inelastic scatterings.

Hence, it is indeed possible to state without hesitation that 
most of contemporary "experimental results" elaborated via the conventional scattering 
theory can be at best  qualified as being "experimental beliefs" and they will remain so 
until a covering scattering theory is worked out.
\vskip0.30cm

{\it IRREVERSIBILITY: } Always for the studious intent of extending the arena of applicability 
of SR beyond that 
od its original conception, it was generally believed during the 20-th century that 
irreversibility is a macroscopic phenomenon because, according to such a theology, when 
macroscopic objects are reduced to their elementary point-like constituents, irreversibility 
"disappears."

On serious scientific grounds, there are theorems (provable by graduate students in 
physics) establishing that a macroscopic irreversible event cannot be consistently reduced 
to a finite number of particles all in reversible conditions (and vice versa). Hence, irreversibility 
originates at the ultimate level of nature, that of elementary particles.

Since all known potentials are reversible, the only possible or otherwise scientific 
origin of irreversibility is provided precisely by the contact, nonlocal and nonpotential interactions 
considered above.

But the entire mathematical and physical structure of SR is reversible in time, as well known. 
Consequently, any belief of the exact characters of SR for irreversible processes, 
such as deep inelastic scatterings to mention only one, can only be claimed to be purely  
"theological," since the only credible scientific issue is the search for a suitable 
covering of SR that is as irreversible in mathematical and physical structure as the 
phenomena to be represented.

At this point, and only following the above technical analysis, we can identify the 
unreassuring implications for society of the theological condition of particle physics. 
On one side, society has a compelling need  for new clean energies, 
as now denounced by various heads of states and scientists alike. On the other side, 
all energy releasing processes are irreversible, from fire discovered since 
the birth of human civilization to the most advanced nuclear energies. 

Hence, the continued insistence in reducing the entire universe to abstractions compatible 
with SR is much more than a  "theological" posture since it 
constitutes a real threat to society.
\vskip0.50cm

\noindent {\bf LIMITATIONS OF QUANTUM MECHANICS}

\noindent It is generally believed that Quantum Mechanics (QM) is universally valid for all possible conditions 
existing in the universe. This belief too is more than purely "theological" because it 
requires the raising of serious issues of scientific ethics and accountability  
for the very survival of the human society.

There is no doubt that, for the conditions under which particles can be credibly abstracted 
to be point-like, QM is indeed exactly valid. This is the case for the 
hydrogen atom where QM achieved majestic results with the exact representation of 
all spectral lines from first unadulterated axioms 
or the construction of particle accelerators 
via relativistic QM that do indeed work, in which cases particles can  be 
approximated as being point-like while moving in vacuum. Despite these historical results, QM has 
its own clear limitations, such as:
\vskip0.30cm

{\it ATOMIC PHYSICS:} In the transition from the hydrogen atom to the helium, there exist clear 
deviations of the prediction of QM from experimental data; these deviations increase with the 
number of atomic electrons; and the deviations become embarrassing when passing to a QM 
description of heavy atoms such as the zirconium.

The only possible reason for the insufficiency is the presence in systems with many electrons   
of interactions beyond the representational capabilities of QM, which interactions can only 
be of nonpotential type, that is, not derivable from a Lagrangian or a Hamiltonian.

Perhaps more dramatic has been the failure by QM  to achieve any quantitative representation 
of the spectral emission of the Sun in over one century of studies.

Under the above conditions, the belief of the universal validity of QM is not only "theological"  
but actually dangerous to society, since the only scientific issue is the 
search for a covering - broadening of QM capable of achieving a more credible representation 
of nature.
\vskip0.30cm

{\it SUPERCONDUCTIVITY: } The current status of superconductivity can be compared to the status of atomic physics 
prior to the quantum description of the atomic structure. In fact, QM has 
indeed permitted a first description of superconductivity, 
but via an ensemble of Cooper pairs considered as point-like without a quantitative description of their 
structure (because identical electrons are predicted by QM to repel each other, and are certainly not 
predicted to enter into their deep correlation - bond as need for Cooper pairs).

Anybody who believes that superconductivity can reach truly basic advances without a quantitative 
description of the structure of its main element, the Cooper pair (hence, 
 without surpassing QM via a suitable 
covering theory), rather than  doing "theological  speculations," raises serious issues of scientific ethics and accountability.  
\vskip0.30cm

{\it CHEMISTRY:}  As stressed above, QM is exactly valid for the structure of "one"  hydrogen atom. However, 
the same discipline is no longer exactly valid for the structure of "two" hydrogen atoms bonded into 
the hydrogen molecule because of the historical inability to represent a residual 2 \% of the binding 
energy from unadulterated first principles. 

Attempts to bypass this insufficiency have been conducted  
via the so-called "screening of the Coulomb law" and then the claim that quantum mechanics and 
chemistry are still valid. In reality, the quantum of energy is solely possible for the 
unadulterated Coulomb law and it is impossible for any of its various screenings, for which 
no quantized orbit can be defined, thus raising serious issues of scientific 
ethics and accountability for the very use of the name "quantum mechanics and chemistry" 
under the conditions considered.

The above insufficiency of QM should not be surprising because it can be predicted from its  
inability to reach an exact representation of the spectral emissions of the helium, since the 
hydrogen molecule and the helium both have two electrons.

Besides the above basic insufficiency, QM had prohibited a quantitative study of 
molecular bonds because nuclei do not participate in such bonds due to their large 
mutual distance, as a result of which molecular structures originate solely from valence bonds. 
But two electrons are predicted to repel each other according to QM. Hence, the various notions 
of valence are pure nomenclatures deprived of scientific, that is, quantitative content.

QM is also responsible for numerous additional insufficiencies or sheer inconsistencies in 
chemistry. One of them is the prediction that all substances are paramagnetic. This is a necessary 
consequence of the absence of an actual, attractive force between identical valence electrons 
in which case the electron orbitals of different atoms are independent. It then follows 
from quantum electrodynamics that the applicability of an external magnetic field  causes 
the polarization of the orbitals of all the atoms of any chemical species, resulting in the indicated 
prediction of universal paramagnetic character in dramatic disagreement with reality.
\vskip0.30cm

{\it NUCLEAR PHYSICS.]:} There is no doubt that QM has achieved historical results in nuclear physics 
or that nuclear power plants conceived and constructed via quantum laws do indeed work. 

However, any belief that QM is exactly valid in nuclear physis is asocial because of an excessive 
number of unresolved basic insufficiencies.

As an example, following one century of efforts,  QM failed to achieve an exact 
representation of the basic data of the lightest nucleus, the deuteron, since it still misses 
1\% of the magnetic moment following all possible relativistic corrections; QM has been unable 
to represent the spin 1 of the deuteron ground state (sice quantum axioms predict that the ground 
state of two particles with spin 1/2 must be zero); there has been no  credible 
explanation of the stability of the neutron when a constituent of the deuteron; etc.

In passing to heavier nuclei, the insufficiencies of QM can only be qualified as  embarrassing 
thus demanding the construction  of a covering theory more suitable  to 
represent physical conditions existing in the nuclear structure, 
since they are dramatically different than 
the conditions of the original conception of the theory, the planetary structure of the 
atom.

Without entering into additional technical insufficiency, any belief that QM is exactly valid in 
nuclear physics can be easily  disqualified as being asocial, 
let alone "theological,"  from the incontrovertible evidence 
that {\it nuclei do not have nuclei,} namely, nuclei do not have a Keplerian structure. 
Consequently,  the basic symmetries of QM, the 
Galilean and Poincare' symmetries, cannot possibly be exact for nuclei since they  notoriously apply solely to Keplerian or planetary  structures.

The social implications of the "theological" belief of the exact character of QM in nuclear 
physics are rather serious. Recall that, as it is the case for SR, QM is also strictly reversible in 
time. Yet all energy releasing nuclear processes are irreversible. 
Consequently, the restriction of all research 
in nuclear energies to comply with QM, that has been rigidly imposed by academia and governments 
alike during the entire 20-th century, has perhaps done a damage to society of potentially 
historical proportion.
\vskip0.30cm

{\it PARTICLE PHYSICS:}  There is no doubt that contemporary particle physics is a synonym for "quantum 
theology" with a rather severe price to pay by mankind for the increasingly cataclysmic 
climatic conditions of our planet combined with the virtually complete lack by particle 
physics of even attempting their solution.

Following the historical results for the structure of the hydrogen atom, the applicability of QM 
has been extended to all possible conditions of particles with only rare  critical analyses. However, 
this "applicability" has been achieved by introducing free parameter of generally 
unknown physical origin, using them to achieve the fit of experimental data, and then claim that 
QM is exact. 

The most embarassing case of asocial particle theology occurs for the Bose-Einstein correlation whose fit 
of experimental data via QM has requested the two-points correlation function to have four free 
parameters (called the "chaoticity" parameters) to  
claim that relativistic quantum mechanics is exactly valid in the field. The problem for the 
supporter of such a view is that four parameters are prohibited by the quantum axiom of expectation 
values (because Hermitean operators in this case have only two diagonal elements, 
thus allowing only two, rather than four parameters). Consequently, rather than supporting QM, 
the chaoticity parameters are a direct measurement of the deviation 
of QM from the physical reality of the Bose-EInstein correlation.

A fully similar situation occurs in various other particle events in which experimental fits are 
reached via {\it ad hoc} parameters and experimental results are claimed while in reality we 
merely have a scientific religion.
\vskip0.30cm

{\it NEUTRON SYNTHESIS AND DECAY:} As originally conceived by Rutherford in 1920 and verified by Chadwick in 1932, 
the neutron is synthesized in the interior of stars from the hydrogen atoms, namely, from a proton 
and an electron.

While, on one side, all physicists admit this incontrovertible reality (because stars initiate their lives as being solely composed of hydrogen, thus synthesizingh first the neutron), on the other side, very few particle physicists admit that QM  
is completely unable to represent such a synthesis  to such an extent  that voids the 
use of  {\it ad hoc} parameters because:

1)  The familiar reaction $p + e \rightarrow n + \nu$ violates the conservation of the
energy unless the proton and the electron have a kinetic 
energy of at least 0.78 MeV (in which case there is no energy left for the neutrino). In 
fact,  the neutron rest energy (939.56 MeV) is 0.78 MeV  bigger than the sum of
the rest energies of the proton and the electron (938.78  MeV). 

2) Recall that all QM bound states (such those for nuclei, atoms and molecules) have a 
"negative" binding energy. The synthesis of the neutron requires instead a "positive" binding-like 
energy. Few physicists know that Schroedinger's equation becomes physically inconsistent for 
the latter systems, as the reader is encouraged to 
verify by trying to solve any quantum bound state in which the conventional negative binding energy 
is turned into a positive value. When this inconsistency is attempted to bypass by assuming that 
the proton and the electron have a relative energy of at least 0.78 MeV, 
the synthesis is prohibited because at that value the p-e cross section is extremely small. 

3) Assuming that the above basic problems can be solved along preferred orthodox lines, QM cannot 
provide a  representation of the meanlife, charge radius, and anomalous magnetic  moment of the neutron.

Rather than admitting the need of a covering of QM capable of a quantitative 
representation of one of the most fundamental events in the universe, a widespread posture in support of 
the quantum theology is that of ignoring the synthesis altogether. In the scientific reality, 
	QM is inapplicable for the synthesis not only of the neutron, but also of 
all hadrons at large.

The denial of the inapplicability of QM for the neutron synthesis can cause 
a damage to society  of potentially historical proportions because the neutron is the 
biggest and most inextinguishable reservoir of clean energy available to mankind since it decays 
spontaneously (when isolated) into a very energetic and easily trapped electron plus the innocuous 
neutrino (assuming it exists). A serious scientific resolution of the synthesis of the neutron is 
expected to lead to means for its stimulated decay in a selected class of light natural stable isotopes into 
equally light, natural and stable isotopes with lower mass, thus  resulting in a much needed 
clean energy due to the lack of harmful radiations as well as the absence of radioactive waste.

Besides, the achievement of a serious knowledge on the possible 
laboratory synthesis and 
stimulated decay of the neutron  could have far reaching implications for society, 
such as permitting conceivable means to stimulated the decay of radioactive nuclear 
waste, thus rendering environmentally acceptable nuclear power plants as currently 
available. How can any physicist in the field  oppose or otherwise ignore efforts at 
surpassing QM via a suitable covering theory for the study of environmentally acceptable 
energies and still expect respect?

The author hopes to be remembered for his irreconcilable disagreement with academic colleagues and 
laboratory directors on the fact that QM and SR "cannot" be exactly valid for the 
synthesis and, therefore, the structure of hadrons (thus demanding specific experiments) for numerous reasons, such as: the impossibility 
of the exact validity of local-differential theories within hyperdense media requiring nonlocal-integral 
effects; the inability to fit experimental data from first principles as done for the structure of 
the hydrogen atom; the evidence that hadrons have no nuclei, thus prohibiting the exact validity of 
fundamental spacetime  symmetries such as the Galileo and the Poincare' symmetries; and other reasons well known in the technical literature.

One of several resolutory experiments the author has recommended over decades to SLAC, FERMILAB, CERN and other particle laboratories is the measure of the behavior of the meanlife of unstable particles with energy, which behavior is expected to deviate from Einsteinian predictions in view of internal nonlocal and non-Hamiltonian effects. Regrettably, this and other resolutory tests have been  ignored in favor of experiments known to be aligned with pre-existing theories following the use of {\it ad hoc} parameters, thus perpetrating the current status of "theological speculations," rather the conduction of research that will resist the test of time.
\vskip0.50cm

\noindent {\bf INSUFFICIENCIES OF THE STANDARD MODEL FOR QUARK CONJECTURES}

\noindent Following the above indicated conception of the neutron by Rutherford, and its verification by Chadwick, 
Heisenberg introduced the SU(2) isospin symmetry for a geometrical unified treatment of protons and 
neutrons, which symmetries was subsequently extended to SU(3), to reach the current formulation 
of the standard model. 

As a result of historical achievement, we believe that 
 the standard model has indeed achieved the final Mendeleev-type classification 
of particles into families. In fact, all fashinating predictions of new particles, 
subsequently confirmed by clear experiments, can be all reduced to the 
primitive classification capabilities of the theory.

However, the hypothesis that quarks 
are physical particles in our spacetime 
is a pure Barton's "theology" because it 
has remained afflicted by a plethora of fundamental unresolved (as well as generally 
unspoken) insufficiencies, such as:

1) According to the standard model, at the time of the synthesis of the neutron  
the proton and the electron literally "disappear" from the universe to be 
replaced by hypothetical quarks as neutron constituents. Moreover, at the time of the 
neutron spontaneous decay, the proton and the electron literally "reappear" again. 
Both these views are  repugnant to  
 scientific reason, because the proton and the electron are the only stable 
 massive  particles clearly established so far and, as such, they simply cannot "disappear" and 
 then "reappear" because so desired by quark supporters.   The only plausible 
 hypothesis for the neutron synthesis p + e => n + v 
  is that the proton and the electron are actual physical 
 constituents of the neutron as originally conjectured by Rutherford, although the latter 
 view requires the adaptation of the theory to physical reality, rather than the opposite 
 attitude implemented by quark "theologies".

2) When interpreted as physical particles in our spacetime,  quarks cannot experience 
any gravity. As clearly stated by Albert Einstein in his limpid writings, gravity can only 
be defined in our spacetime, while quarks can only be defined in the  mathematical, internal, 
complex valued unitary space with no possible connection to our spacetime (because prohibited 
by the O'Rafearthaigh's theorem). Consequently, physicists who support the hypothesis that 
quarks are the physical constituents of protons and neutrons, thus of all nuclei, should see 
their body levitate due to the absence of gravity.

3) When, again, interpreted as physical particles in our spacetime,  quarks cannot have 
any inertia. In fact, inertia can only be rigorously admitted for the eigenvalues of the 
second order Casimir invariant of the Poincare' symmetry, while quarks and their masses  cannot be defined 
with such a basic spacetime symmetry, as expected to be known by experts to qualify as such. 
Consequently, the idea that quarks have physical masses is pure "theology" deprived of true 
scientific content. In reality, "quark masses" are arbitrary parameters used to fit things.

4) Even assuming that, with unknown scientific manipulations, the above inconsistencies are
resolved, it is known that quark theories have failed to achieve a representation 
of all  
characteristics of hadrons, with catastrophic insufficiencies in the representation of spins,
magnetic moments, mean lives, charge radii and other basic features of hadrons. The sole need to confine quarks due to the lack of their detection should be sufficient for their dismissal as physical particles since a serious  confinement can only be achieved via the by-passing of Heisenberg's uncertaintly principle that always admit a finite probability for quarks to be free. Hence, the very conception of quarks is in conflict with QM axioms (the sole 
possibility to have an identically null probability for quarks to tunnel outside is by rendering incoherent the interior and exterior Hilbert spaces, namely, by assuming a generalzed mechanics in the interior of particles).

5) It is also known by experts that the application of quark conjectures to the structure of nuclei has
multiplied the controversies, while resolving none of them. As an example, the assumption
that quarks are the constituents of protons and neutrons in nuclei has failed to achieve a
representation of the main characteristics of the simplest possible nucleus, the deuteron. In fact, quark
conjectures multiply the limitations of QM to represent the spin 1 of the deuteron (since there are 
problems even in representing the spin or the proton and of the neutron), of 
  the anomalous magnetic moment of the deuteron (because the "theological" quark orbits are 
  too small to allow the needed polarizations), or the 
  stability of the neutron when a deuteron constituent, they are unable to represent the 
charge radius of the deuteron, etc.

The author's view is that quarks are indeed necessary for the standard model, and he uses them routinely 
for calculations in the field, trivially, because quarks are the regular representation of the 
SU(3) symmetry. Yet, after decades of studies in the field, the author has been unable to 
identify truly serious reasons for quarks to be physical particles in our spacetime. At any rate, 
the continuation of claims that quarks are physical particles without a rigorous proof 
that they have gravity goes beyond the level of Barton's "theology" since it raises serious 
problems of scientific ethics and accountability.

On historical grounds, the classification of nuclei, atoms and molecules required two different
models, one for the classification and a separate one for the structure of the 
individual elements of a
given family. Quark theories depart from this historical teaching because of  their 
conception of representing with one single theory both the classification and the 
structure of particles. 

The view advocated is that, quite likely, history will repeat itself. 
The transition from the Mendeleev classification of atoms to the atomic structure 
required a basically new theory, QM. Similarly, the transition from the 
Mendeleev-type classification of particles to the structure of individual particles will 
require a broadening of the basic theory, this time a generalization of QM  
due tg thedramatic differences of the dynamics of particles moving 
in vacuum, as in the atomic structure, to the dynamics of particles moving within 
hyperdense media as in the hadronic structure.

In the final analysis, the  "theology" that quarks are physical particles in our spacetime is, by far, 
the most dangerous for mankind because it prevents even the consideration of the new clean 
energy contained in the neutron, trivially, because according to the standard model the neutron 
constituents cannot be released free. By comparison, if the proton and the electron are the physical 
constituents of the neutron according to Rutherford's conception, the electron can be excited by 
stimulating the neutron's decay with the understanding that QM must be necessarily abandoned 
in favor of a covering theory developed for the synthesis, as stressed earlier.

The reader should not forget that nuclear, atomic and molecular structures have made 
momentous contributions to mankind precisely because their constituents can be produced free. 
By comparison, quark conjectures have made no practical contributions whatever and none 
is even remotely conceivable precisely because of the "theology" that quarks 
are permanently confined.

In short, the  insufficiencies 
of the quark hypothesis are a mere manifestation of Richter's "theological speculations," not 
on marginal aspects, but on the truly fundamental aspect, the impossibility for 
QM and SR to be exactly valid for the synthesis, structure, scattering and decays of particles.
\vskip0.50cm

\noindent {\bf INSUFFICIENCIES OF THE STANDARD MODEL FOR NEUTRINO CONJECTURES}

\noindent Following Pauli's indication that the synthesis of the neutron from a proton and an electron 
according to QM misses spin 1/2, and Fermi's hypothesis of the neutrino (meaning "little 
neutron" in Italian), neutrino physics is today part of the standard model.

Despite clearly historical studies, neutrino physics is indeed a clear example of Barton's 
"theological speculations" because it  has remained afflicted by  fundamental, 
unresolved (as well unreassuringly unspoken), theoretical and experimental problems, such as:

1) Neutrino physics is based on an excessive number of individually unverifiable assumptions. 
In fact, the original hypothesis of one massless neutrino, was replaced by the sequential 
hypotheses that: there exist three different neutrinos and their antiparticles; neutrinos 
have masses;   neutrino  masses are different; neutrinos  ``oscillate''; with additional 
hypotheses expected due to the known insufficiencies  of all preceding ones.

2) According to the standard model, said various neutrinos  can traverse very large hyperdense 
media (such as entire stars) without any collision while being massive particles carrying 
energy in our spacetime. This view is outside scientific reason.

3) As indicated earlier, the synthesis of the neutron $p + e \rightarrow n + \nu$ misses 0.78 MeV. 
In the event the proton and the n eutron have the relative energy of 0.78 MeV, there is no energy left 
for the neutrino and, in any case, the synthesis is not possible due to the virtually null cross section 
of protons and electrons at said energy.

4) Calculations on the ``bell shaped''  form of the energy of the electron in nuclear 
beta decays show that no energy  appears 
to be left for the neutrino, provided hat nuclei  are represented in their actual extended size. In fact,    
the Coulomb interaction between an  extended nucleus and the emitted electron varies
with the direction of the beta emission, with maximal (minimal) value for radial (tangential) 
emissions,  the 'missing energy'' being apparently absorbed by the nucleus.

5) Neutrino experiments are perhaps more controversial than theoretical studies because: 
the number of events used as "experimental evidence" for the existence of neutrinos  is 
excessively small over an extremely large number of events, thus preventing acceptance by 
the physics community at large; experimental data are elaborated with a theory crucially 
dependent on the existence of the neutrinos, in which case the "experimental results" are 
expected to depend on the theoretical assumptions; the theory contains an excessive number 
of parameters (such as the different neutrinos masses and others) essentially capable to 
achiever any desired fit; some of the recent "neutrino detectors" contain radioactive isotopes 
that could themselves trigger the very few selected events; and other reasons.

The author fails to understand an argument often voiced as "evidence" for the existence of the 
neutrino, the experimental evidence on the conservation of the leptonic number. In fact, such an 
argument {\it de facto} implies that the violation of parity in weak interactions is "evidence" for the 
existence of another yet unknown particle.

In reality, the lack of existence of the neutrino as physical particle in our spacetime can stimulate momentous advances, such as conceivable new communications propagating via 
longitudinal impulses through the ether at a speed predicted to be a multiple of that 
of the conventional transversal electromagnetic waves, thus providing hopes that 
mankind may one day initiate interstellar communications for which light signals can only be compared to smoke signals during the early human civilization. Yet, the price to pay 
for these momentous advances  is the abandonment 
of the religious "theology" of the universal validity of SR for all conditions 
existing in the universe.

We believe that all the above insufficiencies are just a manifestation of the truly basic one, 
namely, Richter's "theological speculations" not referred to tangential aspects. but to the 
inapplicability of QM and SR for conditions outside those of their original conception, thus being 
inapplicable as exact disciplines, rather than violated.
\vskip0.50cm

\noindent {\bf INCONSISTENCIES OF GENERAL RELATIVITY AND ASTROPHYSICS}

\noindent While QM and SR have their own arena of exact validity outside which they remain valid as 
an approximation of reality, General  Relativity (GR) is known to be, by far, the most 
controversial scientific "theology" in history because of an excessive number of structural inconsistencies in its conception, such as:

1) We have all been teaching at the graduate school of physics that analytic theories without a Hamiltonian 
cannot represent any dynamics, trivially, because of the impossibility of 
formulating any meaningful time evolution. 
GR has indeed an identically null Hamiltonian and, consequently, any attempt at 
dynamical treatments, such as the evolution of planets, is a pure "theology" deprived of 
serious or otherwise credible scientific content.

2) Assuming that the above basic inconsistency can be somewhat bypassed, GR is noncanonical at 
the classical  level and nonunitary at the operator level. it is today known that these theories 
verify the "theorems of catastrophic inconsistencies." In fact. GR cannot leave invariant in time the 
basic unit of its geometry, thus being unable to preserve in time the 
numerical values of the units of measurements, with consequential catastrophic collapse of  the 
mathematical structure (due to the loss over time of the base field), as well as of the physical structure 
(inability to admit the same numerical predictions under the same conditions at different times, 
violation of causality, and other inconsistencies inherent in all noncanonical or nonunitary theories).

3) Einstein's field equations $G_{\mu\nu} = 0$ are irreconcilably incompatible with quantum 
electrodynamics because they attempt the reduction of gravity to pure curvature without source while 
quantum electrodynamics requires the presence of a first-order source tensor  
even for neutral systems such as the pi-zero meson. In any case, Einstein's attempt to reduce 
gravity to pure curvature without source is disproved by the forgotten fifth identity of the 
Riemannian geometry, the Freud Identity (that requires the presence of two source tensors).

4) GR is not compatible with SR on numerous counts, such as the presence in GR of a well defined Newtonian 
limit in PPN approximation but the absence of any meaningful Minkowskian limit; the impossibility 
for gravitational conservation laws to admit a relativistic counterpart due to the fact that 
the latter are the generators of a symmetry, the Poincare' symmetry, while GR has no symmetry but 
only covariance.

5) It is today known that the bending of the light near astrophysical bodies, that lead to a 
world wide support of GR as well as to one of Einstein's Nobel prizes,  in reality is due to the 
Newtonian attraction of light, and definitely not to curvature.  Also, curvature is irreconcilably 
incompatible with a main gravitational event, the free fall of bodies along a radial line for which 
the notion of curvature has no sense. Additionally, the notion of curvature does not allow a unique 
representation of experimental data since there are several possible PPN expansions and, 
consequently, several possible numerically different gravitational realities. Should the author keep going?

There is no doubt that the religious fervor in support of a catastrophically inconsistent 
theory such as GR, while systematically ignoring incontrovertible inconsistencies, 
has lead to the biggest scientific "theology" in history, including large 
collateral scientific damage, such as: wasting a river of ink and public money 
in attempting a grand unification inclusive of gravitation that is simply beyond rational science due to 
dramatic structural differences; an additional river of ink and public funds to attempt a reconciliation 
of QM and GR that have also been theological inquiries since a 
first year graduate student in physics can prove that, assuming GR admits a consistent operator 
formulation, quantities that are observable (Hermitean) at the initial time are no longer so 
at a later time for a nonunitary theory (Lopez's lemma).

The implications for astrophysics of the widespread extension of SR to conditions it is clearly 
inapplicable plus the use of a GR afflicted by so vast inconsistencies can only be dubbed as dramatic.

To illustrate the gravity of the condition, recall that the statement "universal 
constancy of the speed of light" is one of the most corrupt statement in science when 
not completed with the words "in vacuum," since the verification has solely occurred in vacuum and it 
is today well established that the speed of light is smaller than that in vacuum for transparent 
media of low density while it is generally bigger than that in vacuum for 
transparent media of high density.

. 
At any rate, the extension of the speed of light as the maximal causal speed for the hyperdense 
media in the interior of stars and quasars is beyond scientific reason since 
light does not even propagate  
within these media. But then, the use of the speed of light "in vacuum" to compute the 
energy equivalence of a "hyperdense" star $E = m c^2$, the conjecture of 
dark matter and all that can only be dubbed as "wild theologies" since recently calculated 
maximal causal speeds within the hyperdense interior of stars and quasars are  
so vastly bigger than that in vacuum  to void any need for dark matter.

Similarly, the most plausible explanation not only for the expansion of the universe, 
but also for its acceleration, is that the universe is made up of 
matter and antimatter galaxies under mutual, continuous gravitational repulsion.

This old cosmological conception has been systematically discredited for about one century 
on religious grounds 
that both SR and GR do not admit antigravity. The point where 
serious science turns into a "wild theology" mandating senatorial investigations when 
perpetrated under governmental support is that both the SR an d GR cannot even represent 
antimatter, as recalled earlier.

The "theological" difficulty for new vistas in astrophysics is that they mandate the 
abandonment of SR and GR  for  suitable covering theories more plausible for the interior of matter 
stars and different covering theories for antimatter stars.
\vskip0.50cm

\noindent {\bf THE ONGOING SCIENTIFIC OBSCURANTISM}

\noindent In the 1970s, the author received research support from the United States Air 
Force Office of Scientific Research (USAFOSR). One day, in thge mid 1970s,  the author received a phone call 
from an officer of the USAFOSR informing him that the U. S. Military had decided to terminate 
support for academic research. In fact, soon thereafter funding of academic research was passed to ERDA 
that subsequently became the U. S. Department of Energy (DOE).

At that time, the author was rather naive and asked: "But how can 
the American Military remain strong if funding of academic research is ended?" at which the USAFOSR officer 
promptly replied "Because we cannot allow the security of the United States be hostage to  
pet theories preferred by professors at leading academic institutions."

Following the passage of three decades,, nobody can credibly deny that the U. S. 
Military has  made scientific and technological advances beyond our imagination, while  
 academic research has solely seen "theological 
speculations" such as neutrino or quark conjectures 
 and the like, without any truly basic and/or fundamental advance. 

The reason for the transparent disparity is precisely that stated by the USAFOSR officer, namely, that 
military research has advanced because without any religious attachment to preferred "pet theories," while 
no possible basic advance for academic research could be predicted since the mid 1970s, and none actually  
occurred, because of the systematic restriction of all research funding and academic 
 positions to be "hostage" of preferred "pet theories."

Few academicians know that a number of U. S. corporations are now following the example of the 
U. S.Military, namely, they avoid disclosing to academia 
 the most advanced part of their R\&D. In fact, the author 
has conducted advanced corporate research under contractual obligations not to disclose it 
to academia because of expected abuses of academic authority in the protection 
of preferred "pet theories" causing the termination of corporate research funding.

All these and other unreassuring occurrences are 
a manifestation of the insufficiency of Burton Richter's denunciation of the theological nature 
of contemporary research because, even though deserving full respect and appreciation by any 
responsible scientist or citizen, such a denunciation 
did not have sufficient depth to identify the origin of the 
problem in the religious dominance of preferred "pet theories" now merely belonging to 
the past century.

After some 50 years of research, the author feels obliged to denounce the ongoing existence of a 
scientific obscurantism of such a dimension to dwarf by comparison the scientific obscurantism 
imposed by the Vatican during Galileo's times. 

Even though fully deplorable, the 
origin of the latter obscurantism can be seen in actual religious issues of the time, 
while Galileo's ideas 
eventually proved to have do damaging effect to the Vatican.
By comparison, the origin of the ongoing scientific obscurantism can be seen in very large 
organized interests in pre-existing theories of such a dimension to render lilliputian the religious 
interests during Galileo's time.

Above all, the obscurantism imposed by the Vatican had no impact 
on the people of Galileo's times since it dealt with religious dogmas. 
By comparison, the ongoing scientific obscurantism is having a potentially 
devastating impact on our planet because, 
as shown during these comments,  the solution of the increasingly cataclysmic climactic changes 
is known not to be admitted by quantum mechanics and special relativity but to require their 
surpassing via suitable covering theories. 

The only possibility of avoiding a condemnation by posterity of historical proportion is that 
the U. S., British, German, French and other physical societies implement a serious scientific 
democracy for qualified inquiries in which research and publications along 
 "pet theories" do indeed continue but, 
jointly, physical societies halt their current practices of 
dismissing basic advances via attempted discrdeditations and abuse of authority, rather than credible 
technical arguments. As a matter of fact, to really serve society in a moment of need, 
physical societies should have the opposite posture, that of providing priority to 
much needed basic advances and relegate studies along preferred "pet theories" for what they are, 
"theological speculations" outside contemporary real science.

In short, it is time for the physics community to come to its senses and admit that "basic research" primarily refers to 
the laborious effort of trial and error, not toward tangential issues of 
marginal relevance on pre-existing doctrines, 
but toward basic advanced beyond pre-existing doctrine. After all, the rather 
widespread "theological" belief that quantum mechanics, special and general relativity and the 
standard model are of final character for all events in the universe is  
strictly amoral, ascientific and asocial.
\vskip0.50cm

\noindent {\bf REFERENCES}

\noindent The literature underlying the studies touched in the above comments  is so vast to prevent discriminatory 
partial lists. Serious scholar may consult the 90 pages long General Bibliography available 
at www.i-b-r.org/Hadronic-Mechanics.htm
The same web site presents conceivable covering theories for the solution of the 
insufficiencies and/or inconsistencies herein considered. Additional specific studies can be 
located in the www.arxiv.org by searching papers under the author's name, and inspecting the quoted 
references.

\vskip0.50cm

\noindent  {\large {\bf Ruggero Maria Santilli}}

\noindent Carignano (Torino), Italy

\noindent November 14, 2006 
\vskip0.50 cm

\end{document}